\documentclass[12pt]{article}

\catcode`\@=11

\global\arraycolsep=2pt
\usepackage{amsbsy,amssymb,latexsym,amsfonts,amsmath}
\usepackage{graphicx,color}

\allowdisplaybreaks

\begin{document}
\begin{flushright}
\parbox{4.2cm}
{YITP-23-76}
\end{flushright}

\vspace*{0.7cm}

\begin{center}
{ \Large Higher group Weyl Symmetry}
\vspace*{1.5cm}\\
{Yu Nakayama}
\end{center}
\vspace*{1.0cm}
\begin{center}

Yukawa Institute for Theoretical Physics,
Kyoto University, Kitashirakawa Oiwakecho, Sakyo-ku, Kyoto 606-8502, Japan

\vspace{3.8cm}
\end{center}

\begin{abstract}
We study a higher group analog of the Weyl symmetry in four-dimensional quantum field theories. A typical example is that the modified transformation of the 2-form background gauge field replaces the operator-valued Weyl anomaly associated with gauging the 0-form 
 global symmetry. It is analogous to the 2-group global symmetry where the modified transformation of the 2-form background gauge field replaces the operator-valued chiral anomaly associated with gauging the 0-form symmetry.
 The physical origin of the higher group Weyl symmetry is that under the renormalization group flow, the conserved current mixes with the electromagnetic current that couples with the dynamical gauge field. We can express these effects in the local renormalization group equations in a unified manner. 
\end{abstract}

\thispagestyle{empty} 

\setcounter{page}{0}

\newpage

\section{Introduction}
A modern reformulation of symmetries in quantum field theories has led to a novel understanding of the structures, new consequences, and phenomenological predictions (see e.g. \cite{Cordova:2022ruw}\cite{Schafer-Nameki:2023jdn}\cite{Brennan:2023mmt} for recent reviews). The Weyl symmetry (scale or conformal symmetry), however, has resisted such a reformulation so far. Since the scale transformation is intrinsically related to the renormalization group or holography, pursuit in this direction should be a crucial next step.

In this paper, we study a higher group analog of the Weyl symmetry in four-dimensional quantum field theories. A typical example is that the modified transformation of the 2-form background gauge field replaces the operator-valued Weyl anomaly associated with gauging the 0-form global symmetry. It is analogous to the 2-group global symmetry where the modified transformation of the 2-form background gauge field replaces the operator-valued chiral anomaly associated with gauging the 0-form symmetry \cite{Cordova:2018cvg}\cite{Benini:2018reh}.

The higher group global symmetries\footnote{For earlier works see e.g. \cite{Kapustin:2013qsa}\cite{Kapustin:2013uxa}\cite{Gukov:2013zka}\cite{Sharpe:2015mja}\cite{Tachikawa:2017gyf}.  For other applications in four-dimensional physics, see e.g. \cite{Hsin:2020nts}\cite{Iqbal:2020lrt}\cite{Lee:2021crt}\cite{Bhardwaj:2021wif}\cite{Damia:2022seq}\cite{Kan:2023yhz}.} have several applications in particle physics, such as the energy scale of the spontaneous breaking or emergence of the (higher-form) global symmetries \cite{Tanizaki:2019rbk}\cite{Hidaka:2020iaz}\cite{Hidaka:2020izy}\cite{Brennan:2020ehu}\cite{Brauner:2020rtz}\cite{Nakajima:2022feg}\cite{Nakajima:2022jxg}\cite{Cordova:2022qtz}. In particular, it has been much studied in the context of axion dynamics. In our story, in addition to the axion, a dilaton, associated with the spontaneous breaking of the Weyl or conformal symmetry, will appear.

At the beginning of this paper, we have addressed the importance of the reformulation of the Weyl symmetry in the modern viewpoint. This is even more so if we try to incorporate supersymmetry. The supersymmetry relates the R-current anomaly with the Weyl anomaly. Indeed, we will argue that the higher group global symmetry and the higher group Weyl symmetry may have the same origin in supersymmetric field theories.

The organization of the paper is as follows. In section 2, we introduce the concept of higher group Weyl symmetry in drawing an analogy with the higher group global symmetry. We show that these effects can be expressed in the local renormalization group equations in a unified manner.  In section 3, we give further discussions on the property of the higher group Weyl symmetry.

\section{Higher group Weyl symmetry}

\subsection{2-group global symmetry}
Let us present an example of the 2-group global symmetry, which is to be contrasted with the 2-group Weyl symmetry that we will introduce in section 2.2. Consider a quantum field theory with the $U(1)^{(0)}_A \times U(1)^{(0)}_B \times U(1)^{(0)}_C$ global symmetry in four dimensions. The superscript $(0)$ means that it is a 0-form symmetry that acts on point-like local operators. The subscript $A, B, C$ distinguish the $U(1)$ symmetry while it is possible to identify two of them (e.g. $A$ and $B$), and it is often the case in the examples studied in the literature \cite{Cordova:2018cvg}, but for our purpose to draw an analogy in section 2.2, we start with three global symmetries.

Let us further assume that there is a mixed 't Hooft anomaly $\kappa_{ABC}$ among them. We will gauge the  $U(1)^{(0)}_C$ symmetry in a moment, and we choose the local counter-term\footnote{We may add $\int d^4x \epsilon^{\mu\nu\rho\sigma}(c_a A_\mu^B A_\nu^C F_{\rho\sigma}^A + c_b A_\mu^C A_\nu^A F_{\rho\sigma}^B + c_c A_\mu^A A_\nu^B F_{\rho\sigma}^C) $ as local counter-terms, two terms out of three being independent. What we mean by the anomaly hereafter includes the particular shift chosen at this point.} so that the associated current  $J^\mu_{C}$ is conserved under the presence of the background gauge field for the $U(1)^{(0)}_A \times U(1)^{(0)}_B$ symmetry. In addition, we assume that there is no gauge anomaly (i.e. $\kappa_{CCC}$) so that we can gauge the $U(1)^{(0)}_C$ symmetry with no obstruction.

We now replace the background gauge field $A^\mu_C$ with the dynamical gauge field $a^\mu_C$ and add suitable kinetic terms (e.g. $\frac{1}{4e_C^2} g^{\mu\rho} g^{\nu\sigma} f_{\mu\nu}^C f_{\rho\sigma}^C$, where $f_{\mu\nu} = \partial_\mu A^C_\nu-\partial_\nu A_\mu^C$). Let us see what happens to the global symmetry $U(1)^{(0)}_A$. The original 't Hooft anomaly from the $\kappa_{ABC}$ now becomes the ``operator-valued anomaly"
\begin{align}
\partial_\mu J^\mu_A = \kappa_{ABC} (\frac{\epsilon^{\mu\nu\rho\sigma}}{2} F_{\mu\nu}^B f_{\rho\sigma}^{C}) + \cdots \ .  
\end{align}
It is operator-valued because $f_{\rho\sigma}^{C}$ is dynamical (while $F_{\mu\nu}^B$ is a background field).
We could have the other 't Hooft anomaly (e.g. from non-zero $\kappa_{ABB}$) hidden in $\cdots$, but for our presentation purpose, we neglect them for now.\footnote{In addition, the presence of $\kappa_{ACC}$ (i.e. ABJ anomaly) implies that the $U(1)^{(0)}_A$ 
 transformation is no longer an invertible symmetry. We could still have a non-invertible symmetry by augmenting topological degrees of freedom \cite{Choi:2022jqy}\cite{Cordova:2022ieu}.}

The idea of the 2-group global symmetry is to rewrite the  ``operator-valued anomaly" in the form of the modified transformation of the background $2$-form gauge field. We first note that the dynamical gauge field for the $U(1)^{(0)}_C$ symmetry induces the 1-form global symmetry $U(1)^{(1)}_C$ associated with the Bianchi identity $\partial_\mu (\epsilon^{\mu\nu\rho\sigma} f_{\rho\sigma}^C )= 0$. Then we can introduce the background 2-form gauge field $A_{\mu\nu}^C$ with the coupling $\frac{\epsilon^{\mu\nu\rho\sigma}}{2} A_{\mu\nu}^C f_{\rho\sigma}^C $. We now recognize that the ``operator-valued anomaly" can be expressed as the modified gauge transformation of the 2-form background gauge field with the 2-group structure:
\begin{align}
\delta A^A_\mu &= \partial_\mu \lambda^A  \cr 
\delta A_{\mu\nu}^C & = \kappa_{ABC} \lambda^A F_{\mu\nu}^B \ , \label{2global}
\end{align}
where $\lambda^A$ is the gauge parameter.
A similar structure can be obtained by replacing $A$ with $B$.

At this point, it may be worth commenting on the quantization of the coefficient $\kappa_{ABC}$  \cite{Cordova:2018cvg}. If we assume the compactness of the gauge group, i.e. $\int d\lambda_A = 2\pi \mathbb{Z}$, the ambiguity of the gauge parameter $\lambda_A \sim \lambda_A + 2\pi$ induces the ambiguity of $A^{\mu\nu}_C$. 
Once we assume the quantization of the flux $\int F_B = 2\pi \mathbb{Z}$, it is consistent only when $\kappa_{ABC}$ is quantized.
 This is in accord with the group cohomology $H^3(U(1)_A^{(0)}, U(1)_C^{(1)}) = \mathbb{Z}$. 

Finally, note that if we make the background $U(1)_C^{(1)}$ gauge field $A_{\mu\nu}^C$ dynamical, it will remove the $U(1)_C^{(0)}$ gauge field dynamically, leading to the original ungauged theory with the  $U(1)^{(0)}_A \times U(1)^{(0)}_B \times U(1)^{(0)}_C$ global symmetry.

In the above, we have used the terminology that ``the operator-valued anomaly can be expressed by the modified transformation of the background gauge field". We may use the same mechanism to ``cancel" the anomaly by making the background gauge field dynamical and assume the modified shift of the dynamical gauge transformation in the opposite direction. This is sometimes called the Green-Schwartz mechanism.

\subsection{2-group Weyl symmetry}
After having reviewed the 2-group global symmetry from the gauging of the $U(1)^{(0)}_A \times U(1)^{(0)}_B \times U(1)^{(0)}_C$ global symmetry with the mixed 't Hooft anomaly, we now introduce the 2-group Weyl symmetry.  The idea is to replace one of the 0-form global symmetries, say $U(1)_A^{(0)}$, with the Weyl symmetry. The Weyl symmetry acts on the background metric $g_{\mu\nu}$ as $\delta g_{\mu\nu} = 2\delta \sigma g_{\mu\nu}$, where $\delta \sigma$ is space-time dependent. 
The role of the conserved current is then replaced with the trace of the energy-momentum tensor $T^\mu_\mu$. 

To exemplify the situation, let us consider a quantum field theory with the $U(1)_B^{(0)} \times U(1)_C^{(0)}$ global symmetry in four dimensions. We assume that there is a mixed Weyl anomaly $\tau_{BC}$ associated with the background gauge fields. More explicitly, the trace of the energy-momentum tensor has the form
\begin{align}
T^\mu_\mu = \tau_{BC} (g^{\mu \rho} g^{\nu\sigma} F_{\mu\nu}^B F_{\rho\sigma}^C) + \cdots \label{Weyl}
\end{align}
While we could have the other terms in the Weyl anomaly hidden in $\cdots$, we focus on this particular term. We will come back to the physical implications of the other terms toward the end of this subsection.

Let us replace the background gauge field $A_\mu^C$ with the dynamical gauge field $a_\mu^C$ and add suitable kinetic terms (e.g. $\frac{1}{4e_C^2} g^{\mu\rho} g^{\nu\sigma} f_{\mu\nu}^C f_{\rho\sigma}^C$). We should assume that gauge anomaly $\kappa_{CCC}$ vanishes for this purpose. 
The original Weyl anomaly from $\tau_{BC}$ now becomes the ``operator-valued Weyl anomaly"
\begin{align}
T^\mu_\mu = \tau_{BC} (g^{\mu \rho} g^{\nu\sigma} F_{\mu\nu}^B f_{\rho\sigma}^C) + \cdots \label{opWeyl}
\end{align}

In analogy to the 2-group symmetry discussed in section 2.2, the idea of the 2-group Weyl symmetry is to rewrite the  ``operator-valued Weyl anomaly" in the form of the modified transformation of the background 2-form gauge field. This rewriting of the ``operator-valued Weyl anomaly" in the form of the variation of the background field is closely related to the local renormalization group equation as we will see in section 2.3.

The dynamical gauge field for the $U(1)^{(0)}_C$ symmetry induces the 1-form global symmetry $U(1)^{(1)}_C$ associated with the Bianchi-identity $\partial_\mu (\epsilon^{\mu\nu\rho\sigma} f_{\rho\sigma}^C )= 0$. It couples with the background 2-form gauge field $A^{\mu\nu}_C$. We then notice that the ``operator-valued 
 Weyl anomaly" can be expressed as the modified Weyl transformation with the 2-group structure:
\begin{align}
\delta g_{\mu \nu} &= 2\delta \sigma g_{\mu\nu}  \cr 
\delta A_{\mu\nu}^C & = \tau_{BC} \delta \sigma \frac{\epsilon_{\mu\nu}^{\ \ \rho \sigma}}{2}  F_{\rho\sigma}^B \ .  \label{2Weyl}
\end{align}
Since the Weyl transformation induces the non-standard transformation of the 2-form gauge field, it is appropriate to name the structure the 2-group Weyl symmetry.

Since the Weyl transformation is non-compact, there is no quantization condition necessary for $\tau_{AB}$. Indeed, we can come up with models with $\tau_{AB}$ that continuously change if we change parameters in the theory. An example is the AdS/CFT, where $\tau_{AB}$ is given by the bulk gauge coupling constant which can take an arbitrary value. In addition, due to the presence of the gauge kinetic term for the dynamical gauge fields, $\tau_{AB}$ begins to depend on the gauge coupling constant $e_C$.

We have a comment on the term omitted in \eqref{opWeyl}. In unitary theories, there always exist the terms  $\tau_{CC}(g^{\mu \rho} g^{\nu\sigma}F_{\mu\nu}^C F_{\rho\sigma}^C)$ as well as $\tau_{BB}(g^{\mu \rho} g^{\nu\sigma}F_{\mu\nu}^B F_{\rho\sigma}^B)$. They cannot be non-negative in unitary theories because it is given by the current two-point functions. 

Once we make the background field $A^\mu_C$ dynamical, the direct effect from the additional term $\tau_{CC}(g^{\mu \rho} g^{\nu\sigma}f_{\mu\nu}^C f_{\rho\sigma}^C)$ is that it make the gauge coupling constant for the $U(1)^{(0)}_C$ run. This running effect is another ``operator-valued Weyl anomaly". We may rewrite the renormalization group running of the gauge coupling constant by introducing the background space-time dependent gauge coupling constant $g(x) = \frac{1}{4e^2(x)}$ and introduce the additional modified transformation 
\begin{align}
\delta g(x) = \tau_{CC} \delta \sigma \ . 
\end{align}
We might want to say that we have introduced the background ``0-form gauge field" $g(x)$ for the ``$-1$-form symmetry" \cite{Cordova:2019jnf} with the modified Weyl transformation law. In this sense, it could have been called the 0-group Weyl symmetry.

The second, more indirect, effect is that we expect quantum corrections of $\tau$s due to the gauge coupling constant, which runs under the renormalization group. The effect is encoded $g  (= \frac{1}{4e^2})$ dependence on $\tau_{BC}$ (as well as $\tau_{CC}$ and $\tau_{BB}$). Unlike the 2-group global symmetry, the parameters are not quantized and can change when we follow the renormalization group flow.

As in the case of the 2-group global symmetry, if we make $U(1)_C^{(1)}$ gauge field dynamical, it will remove the $U(1)_C^{(0)}$ gauge field dynamically, leading to the original ungauged theory with the  $U(1)^{(0)}_B \times U(1)^{(0)}_C$ global symmetry. One potential difference in the 2-group Weyl symmetry here is that the modified Weyl transformation of the 2-form gauge field makes it more difficult to construct a kinetic term for the 2-form gauge field that is compatible with the Weyl symmetry (although this does not affect the statement above). See the construction of the Weyl invariant kinetic term for the 2-form gauge field in a theory with spontaneously broken Weyl symmetry at the end of section 3.

\subsection{Local renormalization group equation}
We can summarize the 2-group Weyl symmetry (together with the effect of the 0-group Weyl symmetry) in terms of the local renormalization group equation \cite{Osborn:1991gm}\cite{Nakayama:2013is}\cite{Jack:2013sha}\cite{Baume:2014rla}. The local renormalization group equation is obtained by adding local source terms for all the local operators whose correlation functions we would like to study. Or in a modern language, we are gauging all the $-1$ form symmetries by introducing 0-form gauge fields.
They are collectively denoted by $g^I(x)$ and the corresponding 
 (local) renormalization group beta functions are given by $\beta^I$

We express the effect of the local renormalization group as an action on the Weyl transformation of the effective action $\mathcal{W}= -\log \mathcal{Z}$, which is a functional of all the sources. 
If we restrict ourselves to the relevant parts of the local renormalization group equations, it reads
\begin{align}
\int d^4x 
\delta \sigma \left(2g_{\mu\nu}\frac{\delta}{\delta g_{\mu\nu}} + \tau_{BC} \frac{\epsilon_{\mu\nu}^{\ \ \rho\sigma}}{2} F^B_{\rho\sigma} \frac{\delta}{\delta A^C_{\mu\nu}} + \tau_{CC}  \frac{\delta}{\delta g} + \beta^I \frac{\delta}{\delta g^I}  \right) \mathcal{W} = \mathcal{A} \ .
\end{align}
Here, $\mathcal{A}$ is the so-called Weyl anomaly functional:
\begin{align}
\mathcal{A} = \int d^4x \sqrt{g} \left( \delta \sigma \tau_{BB}(g^{\mu \rho} g^{\nu\sigma}F_{\mu\nu}^B F_{\rho\sigma}^B) + \cdots \right) \ ,
\end{align}
where we may have additional various Weyl anomaly terms  such as the Weyl tensor squared $c\mathrm{Weyl}^2$ or the Euler density $a\mathrm{Euler}$ in $\mathcal{A}$ omitted in $\cdots$. The entire structure can be found in the literature \cite{Osborn:1991gm}\cite{Nakayama:2013is}.
As we have already discussed, $\tau$ and $\beta^I$ may depend on the other coupling constant such as $g$ or $g^I$.

The higher group structure of the local renormalization group equation may become manifest by noting that the effective action $\mathcal{W}$ also satisfies the 2-form background gauge symmetry 
\begin{align}
\int d^4x (\partial_\mu \lambda^C_\nu - \partial_\nu \lambda^C_\mu) \frac{\delta}{\delta A^C_{\mu\nu}} \mathcal{W}  = 0 \ ,
\end{align}
and we can regard the second term in the local renormalization group as the additional shift of the 2-form gauge transformation under the Weyl transformation. The effective action also  satisfies the 1-form background gauge symmetry\footnote{If we introduce the source for the operators charged under the 0-form global symmetry, the right-hand side has additional contributions 
 $\int d^4x \lambda^B \delta g^I \frac{\delta}{\delta g^I} \mathcal{W}$, resulting in the (anomalous) Ward-Takahashi identity.}
\begin{align}
\int d^4x \partial_\mu \lambda^B \frac{\delta}{\delta A^B_{\mu}} \mathcal{W} = \mathcal{A}^B \ ,
\end{align}
up to possible 't Hooft anomaly $\mathcal{A}^B$ (given by $\kappa_{BBB}$).

The physical meaning of the 2-group Weyl symmetry can be directly read from the local renormalization group equation. If we vary the local renormalization group equation with $A^B_\mu$, we can study the renormalization group flow of the correlation functions with $J^\mu_B$. The resulting equation tells that under the renormalization group flow, the conserved current $J_\mu^B$ mixes with the dynamical gauge field by the amount $\tau_{BC} \partial^\nu f_{\nu \mu}^C$: schematically $\frac{\partial}{\partial \log\mu} J^B_\mu = \tau_{BC} \partial^\nu f_{\nu \mu}^C = \tau_{BC} J_\mu^C$. Here, in the last line, we have used  the ``Maxwell equation" and we interpret it as the mixing of the global current with the electromagnetic current that couples with $a_\mu^C$. Note that the mixing preserves the conservation of the original global current. 

One may naturally wonder if there is a 1-group Weyl symmetry generated by $\delta \sigma \beta_\mu \frac{\delta}{\delta A_\mu}$. Yes, it exists. It is known as the vector beta function and the physical significance was studied in the literature \cite{Nakayama:2013ssa}.\footnote{As may be obvious at this point, there is no reason we should restrict to $p$-forms. We could consider spinor or tensor background fields with non-trivial Weyl transformation.} Typically it has the form $\beta_\mu = \rho_I D_\mu g^I$. Among others, it encodes the anomalous dimensions of the (non-conserved) vector operators.

\section{Discussions}
To close the paper, we have several discussions on the 2-group Weyl symmetry. 

In section 2, we have drawn an analogy with the 2-group global symmetry and the 2-group Weyl symmetry. In supersymmetric field theories, it turns out that it is more than an analogy; it is directly related. This is because the R-current and the trace of the energy-momentum tensor transform each other under supersymmetry. It is in the same supersymmetry multiple. 

With the superspace formulation, the energy-momentum tensor multiplet ($T_\alpha \dot{\alpha}$ and $T$)  satisfies the superspace conservation law (see e.g. \cite{Komargodski:2010rb})
\begin{align}
\bar{D}^{\dot{\alpha}}T_{\alpha \dot{\alpha}} + \frac{2}{3} D_\alpha T = 0 .
\end{align}
Here, $\theta^2$ component of $T$ is given by $T^\mu_\mu + i \partial^\mu J_\mu^R$ with $J_\mu^R$ being the R-current. 
In our setup, we assume that the trace chiral multiplet $T$ is given by the field strength chiral multiplet $W_\alpha$, which has $F_{\mu\nu} + i\frac{\epsilon_{\mu\nu\rho\sigma}}{2} F^{\rho\sigma}$ in the $\theta$ component, as 
\begin{align}
T = \tau_{BC}W^{B\alpha} W^C_\alpha  + \cdots \ .  
\end{align}
Assuming that $\tau_{BC}$ is real,\footnote{The imaginary part of $\tau_{BC}$ would give CP-violating terms in the chiral anomaly as well as in the Weyl anomaly. See \cite{Nakagawa:2020gqc}\cite{Nakagawa:2021wqh} for more details.} it encodes the R-current anomaly $\partial_\mu J^\mu_R = \tau_{BC} (\frac{\epsilon^{\mu\nu\rho\sigma}}{2} F_{\mu\nu}^B f_{\rho\sigma}^{C}) + \cdots$ as well as the Weyl anomaly  $T^\mu_\mu = \tau_{BC} (g^{\mu\rho}g^{\nu\sigma}F_{\mu\nu}^B f_{\rho\sigma}^{C}) + \cdots$ . 

When we make the gauge vector multiplet $W^C_\alpha$ a dynamical one $w^C_\alpha$, we may add the background two-form gauge fields encoded in a single chiral multiplet with the modified transformation that generates \eqref{2global} and \eqref{2Weyl}. In the superspace notation,  we introduce the background chiral superfield $b^B_\alpha$ with the modified superWeyl transformation $\delta b^B_\alpha = \delta \sigma W^B_\alpha$, and the superspace action is $\int d^4 x d^2\theta \tau_{BC} w^{C \alpha} b^B_\alpha  + \text{c.c}$. 
Accordingly, we obtain the 2-group symmetry and the 2-group Weyl symmetry simultaneously with the same coefficient proportional to $\tau_{BC}$.\footnote{The R-charge does not have to be quantized and it avoids the quantization condition mentioned at the end of section 2.1.}

This implies that if we formulate the 2-group global symmetry of the R-current in the manifestly supersymmetric way, we have to talk about the 2-group Weyl symmetry at the same time.

So far, we have studied the case when the global symmetries are all Abelian. We now investigate the possibility we have non-Abelian 
 global symmetries as a starting point. Suppose we have two 0-form non-Abelian symmetries (say $SU(2)^{(0)}_B \times SU(2)^{(0)}_C)$ with the mixed Weyl anomaly
 \begin{align}
T^\mu_\mu = \tau_{BC} g^{\mu\rho} g^{\nu \sigma} F_{\mu \nu} ^{B a} F_{\rho \sigma} ^{Ca} + \cdots \  , 
 \end{align}
where the summation over $a$ preserves only the diagonal symmetry. 
 
As before, we will replace the background gauge field $F_{\mu\nu}^{Ca}$ with the dynamical one $f_{\mu\nu}^{Ca}$. The operator-valued Weyl anomaly can be replaced with the modified transformation of the two-form background fields $A_{\mu\nu}^{Ca}$ that couples with the non-Abelian field strength $\frac{\epsilon^{\mu\nu}_{\ \ \rho\sigma}}{2}f_{\mu \nu}^{Ca}$. The physical meaning of the effect is that under the renormalization group flow, two $SU(2)$ symmetries mix, and what we mean by gauging one of the $SU(2)$ should change along the renormalization group flow.

This construction has one advantage and one disadvantage. The advantage is that in the Abelian case, the non-zero $\tau_{CC}$, which is inevitable from the unitarity, makes the $U(1)$ gauge kinetic terms run, and the folklore says that there is no non-trivial fixed point 
 for the $U(1)$ gauge coupling \cite{Nakayama:2022vim} (unless we have a monopole, in which case, the 1-form symmetry is broken). In the non-Abelian case, a non-trivial fixed point for the gauge coupling is possible (as in the Bnaks-Zaks fixed point), and we can talk about the renormalization group fixed point after the gauging, where the Weyl symmetry is unbroken without introducing the background 0-form gauge fields (i.e. beta functions). 

The disadvantage is the meaning of the ``2-group global symmetry" structure is obscured. This is because it is not immediately obvious if we can realize the non-Abelian 1-form symmetry in a physically sensible manner \cite{Gaiotto:2014kfa}. We should stress here that there is nothing wrong with introducing the background 1-form field to rewrite the ``operator-valued Weyl anomaly" in the form of the modified Weyl transformation of the background field. The only problem is whether this background field has any meaning  as the (physical) 1-form non-Abliean symmetry.

One may finally consider the situation in which the higher group Weyl symmetry is spontaneously broken. Then as Nambu-Goldstone bosons, a dilation $\varphi$ (for the Weyl symmetry), axion $\chi$ (for the 0-form symmetry), and photon $a_\mu$ (for the 1-form symmetry) appear. With the background gauge fields for the 0-form symmetry $A_\mu$ and the 1-form symmetry $B_{\mu\nu}$, the effective action looks like
\begin{align}
S = \int d^4x  \sqrt{g} \left( e^{-2\varphi} (\partial_\mu \varphi \partial^\mu \varphi + (\partial_\mu  \chi + A_\mu)( \partial^\mu  \chi + A^\mu))+ \frac{1}{4}f_{\mu\nu} f^{\mu\nu} + \frac{\epsilon^{\mu\nu\rho\sigma}}{2} B_{\mu\nu} f_{\rho\sigma}  + \cdots \right) \ .
\end{align}
The operator-valued Weyl anomaly is encoded by the modified transformation of the two-form background gauge field $B_{\mu\nu}$. Since we have the non-linearly realized Weyl symmetry by a shift of the dilaton $\varphi \to \varphi + \delta\sigma$, one may alternatively use the coupling $\varphi f_{\mu\nu} F^{\mu\nu}$ to express (or cancel) the Weyl anomaly. 

As promised at the end of section 2.2, we now present the Weyl invariant kinetic term for the two-form gauge field 
\begin{align}
S = \int d^4x \sqrt{g} e^{2\varphi} (\partial_{[\rho} (B_{\mu\nu]} - \tau \varphi \frac{\epsilon_{\mu\nu]}^{\ \ \kappa\sigma}}{2} F_{\kappa\sigma}))^2 \ .  
\end{align}
This is possible thanks to the dilaton. It includes higher derivative terms for the 1-form gauge field $A_\mu$. Note that if we define the Weyl invariant field strength by $H_{\rho\mu\nu} = \partial_{[\rho} (B_{\mu\nu]} - \tau \varphi \frac{\epsilon_{\mu\nu]}^{\ \ \kappa\sigma}}{2} F_{\kappa\sigma})$, it satisfies the modified Bianchi identity $\partial^\mu (\epsilon_{\mu}^{\ \nu\rho\sigma} H_{\nu\rho\sigma}) = -\tau \partial^\mu \varphi \partial^\nu F_{\mu\nu} $.
  While there is no strong experimental evidence that we find a dilaton in nature, the detailed study of the dilaton effective action may lead to further non-trivial constraints on the renormalization group flow. 

\section*{Acknowledgements}
The author would like to thank Takahiro Tanaka for his contribution to the journal club on higher group global symmetry, where the author learned a lot about the subject.
This work is in part supported by JSPS KAKENHI Grant Number 21K03581.

\end{document}